\documentclass[a4paper,prl,showpacs,preprintnumbers,amsmath,amssymb,twocolumn]{revtex4}
\usepackage{epsf}
\usepackage{amsfonts}
\usepackage{amssymb}
\usepackage{amsthm}
\usepackage{graphicx}

\newcommand{\vsigma}{\mbox{\boldmath $\sigma$}}
\newcommand{\vS}{\mbox{\boldmath $S$}}
\newcommand{\vJ}{\mbox{\boldmath $J$}}
\newcommand{\no}[1]{: \! #1 \! :} 
\newcommand{\cf}[1]{\langle #1 \rangle}
\newcommand{\1}{\openone}
\newcommand{\phdagger}{\mathop{\phantom{\dagger}}}           
\newcommand{\psiop}[1]{\psi^{\phdagger}_{#1}}                
\newcommand{\psidop}[1]{\psi^{\dagger}_{#1}}                 
\newcommand{\aop}[1]{a^{\phdagger}_{#1}}                     
\newcommand{\adop}[1]{a^{\dagger}_{#1}}                      
\newcommand{\rhoop}[1]{\rho^{\phdagger}_{#1}}                

\begin{document}

\title{Measuring Luttinger Liquid Correlations from\\ Charge Fluctuations in a Nanoscale Structure}
\author{Paata Kakashvili and Henrik Johannesson}
\affiliation{ Institute of
  Theoretical Physics, 
 Chalmers University of Technology and G\"oteborg University,
SE-412 96 G\"oteborg, Sweden}

\begin{abstract}

We suggest an experiment to study Luttinger liquid behavior in a one-dimensional nanostructure,
avoiding the usual complications associated with transport measurements. The proposed setup consists  
of a quantum box, biased by a gate voltage, and side-coupled to a quantum wire by a point contact. 
Close to the degeneracy points of the Coulomb blockaded box, and in the presence of a magnetic field
sufficiently strong to spin polarize the electrons, the setup can be described as a Luttinger liquid interacting
with an effective Kondo impurity. Using exact nonperturbative techniques we predict that the differential capacitance 
of the box will exhibit distinctive Luttinger liquid scaling with temperature and gate voltage. 

\end{abstract}


\maketitle

It is theoretically well established that interacting electrons
in one dimension (1D) do not form a Fermi liquid, but rather a composite
$-$ a {\em Luttinger liquid} \cite{Haldane} $-$ where {\em all}
low-lying excitations 
are collective, and separately carry charge and spin.
Despite intense efforts, however, there are very few experiments
that unambiguously point to Luttinger liquid behavior in a real 1D
electron system.
Quantum wires \cite{Wire} and
single-walled carbon nanotubes \cite{CNT} are prime examples of systems
where the electron dynamics is effectively one-dimensional. Still,
interpretations of relevant experimental data based on Luttinger liquid
theory remain controversial.
In most experiments until now, one has measured \emph{transport} properties,
and it has been notoriously difficult to assess the extent to which external 
sources, contacts, impurities, etc., influence the results.

In this Letter, we propose a \emph{non-transport} experiment on a 1D
nanoscale structure which avoids the problems mentioned above.
The system is composed by a 1D quantum box side-coupled
to a single-mode quantum wire via a point contact (Fig. 1),
and could be built from a gated GaAs semiconductor or cleaved
edge overgrowth structure \cite{Wire}. A magnetic field is applied
such that the electrons become spin-polarized. The charging of the box is then
monitored as a function of an applied gate voltage or, alternatively, as
a function of temperature at a fixed voltage bias. Using a simple
model, we show that this setup can be analyzed
in terms of Luttinger liquid theory. We find
that the differential capacitance of the quantum box has a nonanalytic
dependence on temperature and
gate voltage, with a scaling exponent that encodes the electron
correlations of the system.
This fingerprint of Luttinger liquid behavior should be
possible to identify by charge measurements using the recently
developed single-electron transistor electrometer technique \cite{Dot},
given proper choice of parameters and design of the setup.

We take the quantum box to be sufficiently small to exhibit Coulomb
blockade \cite{Averin}, 
but large enough for the electrons in the box to be modeled by a
(confined) Luttinger liquid. 
More precisely, we study the limit $\delta E \ll k_BT_K \ll
e^2/2C^{\phdagger}_{\Sigma}$, where $\delta E$ is the  
average level spacing of the box close to the Fermi level, $T_K$ is
the temperature  
scale at which correlation effects set in (to be defined below), and
$e^2/2C^{\phdagger}_{\Sigma}$ is the charging energy 
of the box (with $C^{\phdagger}_{\Sigma}$ the full capacitance of the box).
$\delta E$ thus serves as a low-energy cutoff restricting the validity
of our analysis \cite{footnote1}. 

The system can be modeled by a Hamiltonian
\begin{equation} \label{Hamiltonian}
H=H_{el}+H_{c}+H_{tun},
\end{equation}
where 
\begin{eqnarray}
H_{el}&=&\sum_{k,\alpha} \epsilon^{\phdagger}_k \adop{k,\alpha} \aop{k,\alpha} + \sum_{q,\alpha,\beta} 
\hat{U}_{\alpha \beta}(q) \rhoop{q,\alpha} \rhoop{-q,\beta}, \label{Hel} \\
H_{c}&=&\frac{Q_{1}^{2}}{2C^{}_{\Sigma}}+\zeta VQ_{1}, \label{Hc} \\
H_{tun}&=&\frac{t}{\ell}\sum_{k,p}(\adop{k,0} \aop{p,1} + h.c.). \label{Htun} 
\end{eqnarray}
Here $\aop{k,\alpha}$ are the electron destruction operators in the wire $(\alpha\!=\!0)$ and
the box $(\alpha\!=\!1)$, with the energy $\epsilon^{\phdagger}_k$ measured from the Fermi level 
$\epsilon^{\phdagger}_F$.
\begin{figure}[!hpb]
\begin{center}
\includegraphics[width=2.6in]{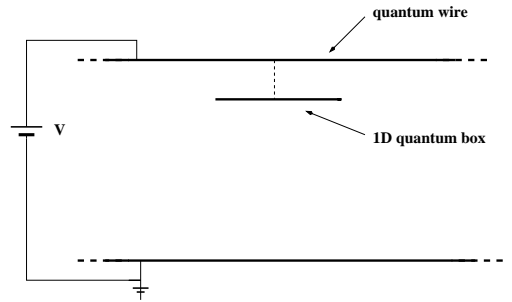}
\caption{Schematic picture of the proposed setup. A 1D quantum box side-coupled
  to a quantum wire via a  point contact. $\bm{V}$ is a gate voltage.}
\label{setup}
\end{center}
\end{figure}
In the interaction term $\rho^{\phdagger}_{q,\alpha}$ are the Fourier components of the corresponding
density operators in the wire and the box, and $\hat{U}_{\alpha \alpha}(q) \, [\hat{U}_{0 1}(q)]$ is the Fourier
transform of the screened interaction potential {\em in} the wire and
the box [{\em between} wire
and box] (with the screening supplied by carriers in nearby gates). 
Since the wire and the box are defined on the same substrate, we shall take $\hat{U}_{0 0}(q) =
\hat{U}_{1 1}(q)$, assuming that their transverse widths are the same. The charging energy
of the box is described by $H_c$, with $Q_{1}$ measuring the surplus charge in the box
w.r.t. the (zero bias) Fermi level, $\zeta$ being a dimensionless parameter which depends 
on the layout of the sample, and $V$ the gate voltage. The last term, $H_{tun}$,
governs the tunneling between the wire and the box, with $t$ the tunneling rate through the
point contact. Note that all effects from the finite size of the 1D box (incl. Coulomb blockade)
are carried by $H_c$, and that in $H_{el}$ and $H_{tun}$ the length $\ell$ of the box for simplicity is taken  
to be the same as that of the extended wire (here assumed to be sufficiently large for additional
charging effects to be ignored). Also note that, while $H_c$ encapsulates only the {\em mean-field}
Coulomb interaction among electrons in the box, the electron-electron interaction
in $H_{el}$ is dynamic and influences the spectrum already for a fixed number of electrons in 
wire and box.

To make progress, we decompose the electron fields 
$\psiop{\alpha}(x)\!\sim \!\int dk \mbox{e}^{ikx} \aop{k,\alpha}$ in left $(\psiop{L, \alpha}(x))$
and right $(\psiop{R, \alpha}(x))$ components (with $x$ the coordinate along the wire), 
expanded about the two Fermi points
$\pm k_F$ of the linearized spectrum.  Keeping only the  ''local'' piece 
$U_{\alpha \beta}(x)\! = \!\hat{U}_{\alpha \beta}(0)\delta(x)$ of the potential, and
setting $\hat{U}_{0 1}(0) = \hat{U}_{0 0}(0)\equiv g$,
$H_{el}$ can be expressed on diagonal Sugawara form \cite{Bos} as
\begin{eqnarray} \label{bulkSugawara}
H_{el} \approx \frac{1}{2 \pi}\int dx\!&\Big
[&\!\frac{v_{c}}{4}(\no{J_{R} J_{R}}+\no{ J_{L}
  J_{L}}) \\ 
\!&+&\!\frac{v_{F}}{3}
(\no{\vJ_{R}\cdot \vJ_{R}}+\no{\vJ_{L}\cdot \vJ_{L}})\Big]\, , \nonumber
\end{eqnarray}
where ''$\approx$'' is a reminder that (\ref{bulkSugawara}) contains the
local part of the interaction only.  
Here $v_{c}=v_{F}(1+4g/v_{F})^{1/2}$, with $v_F$ the Fermi velocity, and
the normal ordering is taken w.r.t. the filled Dirac sea. 
The currents are defined by
\begin{eqnarray}  \label{Currents}
J_{R/L}&\!=\!&\mbox{sinh}\,\vartheta\no{\psidop{R/L,\alpha}\psiop{R/L,\alpha}}
+\,\mbox{cosh}\,\vartheta\no{\psidop{L/R,\alpha}\psiop{L/R,\alpha}} \nonumber \\
\vJ_{R/L}&\!=\!&\frac{1}{2}\no{\psidop{R/L,\alpha}(x) \vsigma^{\phdagger}_{\alpha
  \alpha'} \psiop{R/L,\alpha'}(x)}\, ,  \nonumber
\end{eqnarray}
with $2\vartheta = \mbox{arctanh}(2g/(v_F + 2g))$, $\vsigma$ being the vector 
of Pauli matrices, and the indices $\alpha, \alpha' = 0,1$ summed over. 

One immediately recognizes $H_{el}$ in (\ref{bulkSugawara}) as
a Luttinger liquid Hamiltonian, with dynamically separated charge and ''pseudospin'' currents
$J_{L/R}$ and $\vJ_{L/R}$ respectively \cite{footnote}. Taking into account the boundaries of the box
\cite{MEJ}, as well as the finite range of the screened Coulomb interaction \cite{Schulz}, 
will add more structure to Eq.~(\ref{bulkSugawara}). Also, in a more realistic theory one expects that
$\hat{U}_{0 1} < \hat{U}_{0 0}$, implying that the manifest SU(2) pseudospin symmetry of $H_{el}$ 
in (\ref{bulkSugawara}) gets broken. However, for transparency and ease of notation, we here choose to work with the 
simple theory where $H_{el}$ is represented by (\ref{bulkSugawara}), and return below to discuss the more
general case.

Having built in the Luttinger liquid correlations into the model via (\ref{bulkSugawara}),
we now explore how these influence the charging of the box. Let us first recall that in
a ''classical'' picture the charge in a quantum box biased by a gate voltage $V$
can change only when $V$ is tuned to the discrete values $-ne/2\zeta C^{\phdagger}_{\Sigma}$ (with $n$ an odd integer)
for which the Coulomb blockade is lifted \cite{Averin}. This leads to the celebrated
''Coulomb staircase'' with steps at the {\em degeneracy points} at which 
the charging energy for $(n/2) \pm 1/2$
electrons is the same. This simple picture is modified by quantum charge fluctuations,
enhanced by the coupling of the box to the quantum wire.

To study the fluctuation effects, we probe the system with a gate
voltage close to a degeneracy point, for example
$\zeta V = -e/2C^{\phdagger}_{\Sigma}+u$, with $u \ll e/C^{\phdagger}_{\Sigma}$ (i.e. $u$ is a small voltage 
bias away from the chosen degeneracy point). In the limit of small $t$, we can then truncate the Hilbert space to
the $Q_{1}=0$ and $Q_{1}=e$ states (since in this limit transitions to virtual states of higher energy are
suppressed). Following an exact formulation of Matveev \cite{Mat1}, the resulting
two-level system $H_c + H_{tun}$ in (\ref{Hamiltonian}) can be mapped onto an anisotropic Kondo interaction
\begin{equation}  \label{KondoInteraction}
H_{K}=\frac{J_{\perp}}{2}
\psidop{\mu,  \alpha}(0)
\sigma^j_{\alpha \alpha'}\psiop{\mu',\alpha'}(0) S^j - hS^z \, ,
\end{equation}
where $J_{\perp} = 2t$ and $h=eu$, and where $\vS$ is an additional ''pseudospin'' of magnitude 1/2 that
implements the constraint on the allowed states (with $\vS$ localized at the position $x=0$ of the point contact). 
Note that {\em all} indices in (\ref{KondoInteraction}) ($\mu, \mu'
\!=\! L,R;\, \alpha, \alpha' \!=\! 0,1;\, j\!=\!x,y$) are summed
over. It is here important to realize that  
the presence of backscattering
terms in $H_K$ is due to the fact that the quantum box is {\em side-coupled} to the wire via the point contact. 
This is different from the case of an {\em end-coupled} box, which
supports only forward Kondo scattering 
[11-13].
As it turns out, it is precisely the backscattering in (\ref{KondoInteraction}) that imprints Luttinger liquid characteristics
on the charging of the box, measured by the average $\cf{Q_{1}}$. Its dependence on the gate voltage is given by the 
{\em differential capacitance} $c(u,T)=-[1/(\zeta e^{2})][\partial
\cf{Q_{1}}/\partial V]$, 
which, via the Matveev mapping \cite{Mat1}, gets modeled by an impurity susceptibility $\chi_{imp}(h,T) = \partial \cf{S^z}/\partial h 
\equiv c(u,T)$, describing the response of the local pseudospin to a
''magnetic field'' $h \equiv eu$ at $x=0$. 

The original problem has thus been replaced  
by that of calculating the susceptibility of a (pseudo)spin-1/2 impurity coupled to a Luttinger liquid $H_{el}$
[Eq. (\ref{bulkSugawara})] by an anisotropic Kondo interaction $H_K$
[Eq. (\ref{KondoInteraction})]. 
The presence of backscattering in $H_K$
still makes this a hard problem, however. A perturbative
RG analysis \cite{FN} reveals that the backscattering terms become relevant for interacting electrons, taking the theory to
a nontrivial fixed point. Here we approach the problem 
via a nonperturbative route, exploiting boundary conformal field 
theory (BCFT) \cite{BCFT} to trade the Kondo interaction $H_K$ for a scale invariant boundary condition on the bulk theory $H_{el}$
in (\ref{bulkSugawara}). One can then use BCFT to extract the critical exponents that govern the scaling of $\chi_{imp}$ ({\em alias} the
differential capacitance) for small values of $T$ and $u$ (i.e. close to the fixed point).  

The fixed point describing the {\em isotropic} spin-1/2 Kondo effect in a Luttinger liquid \cite{FN,EK} has been shown to
correspond to a particular selection rule for quantum numbers of the BCFT embedding U(1)$\otimes$U(1)$\otimes$SU(2)$_2$$\otimes$Ising 
\cite{FJ}. Here the two U(1) factors represent the spectra of left-
and right-moving charge excitations, while the
SU(2)$_2$$\otimes$Ising 
block derives from a coset construction of the SU(2)$_1$$\otimes$SU(2)$_1$ left- and right-moving pseudospin excitation spectra 
(with the indices 
labeling the {\em levels} of the corresponding Kac-Moody algebras \cite{Bos}). Given this structure, it is straightforward to verify
that the anisotropy in (\ref{KondoInteraction}) introduces irrelevant operators only (in exact analogy to the Kondo effect
for noninteracting electrons \cite{Anisotropy}). Thus, the fixed point for the present problem is the same as for the isotropic model, and we
can exploit the BCFT scheme developed in Ref.~[\onlinecite{FJ}]. 

Knowing the fixed point allows us to identify the {\em leading} boundary operators that drive the finite-$T$ scaling of $\chi_{imp}$.
Note that, in contrast to the isotropic case in Ref.~[\onlinecite{FJ}], operators that break (pseudo)spin-rotational
invariance are now allowed (by the anisotropy of $H_K$ in (\ref{KondoInteraction})). A systematic search
\cite{KJ} yields two leading operators
$\mathcal{O}^{(1)} = T_s\otimes\1_{Ising}\otimes\1_c$ and $\mathcal{O}^{(2)} =
J^z\otimes\epsilon\otimes\mathcal{O}_c$, with scaling dimensions
$\Delta^{(1)}=2$ and $\Delta^{(2)}=3/2+1/2K_{c}$, respectively.
$K_{c}$ is the usual Luttinger liquid ''charge parameter''
with perturbative expression $K_{c} = (1+4g/v_F)^{-1/2}$
(here allowed to take values in the interval $1/2 \le K_{c} \le 1$),
$T_s$ is the SU(2)$_2$ energy-momentum tensor, $\1$ is the identity operator
in the indexed sector, $J^z$ is the $z$-component of the SU(2)$_2$ pseudospin
current, $\epsilon$ is the Ising energy density, and $\mathcal{O}_c$ is a symmetrized product of U(1) vertex operators
(for details, see Ref.~[\onlinecite{FJ}]).

Given the operators $\mathcal{O}^{(1)}$ and $\mathcal{O}^{(2)}$, the scaling behavior of $\chi_{imp}(T,h\!=\!0)$ can be calculated
via an expansion in their conjugate scaling fields $\lambda_1$ and $\lambda_2$. Passing to a Lagrangian formalism, we write
the partition function as a path integral, treating the (inverse) temperature as an imaginary time. To simplify the calculation
we also replace the local field $h$ in the definition of $\chi_{imp}$ by a uniform field coupling to the pseudospins of all electrons.
This will change the amplitude of the impurity susceptibility \cite{Field}, but since we shall be interested in the
scaling exponents only, this change is immaterial. Using a linked cluster expansion, we can then write
\begin{equation} \label{linkedcluster}
\chi_{imp}(T,0)= \lambda_{1} I[\mathcal{O}^{(1)}_{3}] + \frac{1}{2}\sum_{i=1,2}\lambda_i^2I[\mathcal{O}^{(i)}_{3},\mathcal{O}^{(i)}_{4}]
+ ..., 
\end{equation}
where
\begin{equation}  \label{integraldef}
I[\mathcal{O}_3,...,\mathcal{O}_j]\!\equiv\!\int_{-\infty}^{\infty}\!
\frac{dx_{1}dx_{2}}{4\pi^2\beta}\!\int_{-\beta/2}^{\beta/2}\!d\tau_{1} \dots d\tau_{j}
\cf{J^{z}_{1}J^{z}_{2}\mathcal{O}_{3}...\mathcal{O}_{j}}_{c}, \nonumber
\end{equation}
with $\cf{\dots}_{c}$ a connected n-point function,
and $J^{z}_{k} \equiv J^z(\tau_k, x_k), k=1,2$. Given the boundary operators $\mathcal{O}^{(1,2)}_{j} \equiv
\mathcal{O}^{(1,2)}(\tau_j), j=3,4$, that enter (\ref{linkedcluster}), we use the appropriate operator
product expansions (OPEs) \cite{Bos} to collapse the integrands to products of two-point functions. This
allows us to easily calculate the integrals and we obtain  
(using that $c(T,u\!=\!0)=\chi_{imp}(T,h\!=\!0)$):
\begin{equation}  \label{temperaturescaling}
c(T,u\!=\!0) = A+B[K_{c}]T^{1/K_{c}} + CT^2 + ....,
\end{equation}
with $A$, $B[K_{c}]$ and $C$ constants (where $B[K_{c}]=$const$\{1/K_{c}-1\}$), and where ''....'' indicate subleading corrections.
The short-range electron-electron interaction, encoded by the parameter $K_{c}$, is thus seen to 
induce a {\em nonanalytic term in the differential capacitance}, scaling as $T^{1/K_{c}}$,
while vanishing in the noninteracting limit $(K_{c}=1)$.

Our result in (\ref{temperaturescaling}) predicts a distinct signal of
Luttinger liquid correlations in
the proposed setup. For what temperatures should one expect to see it?
Taking the 1D quantum box to have a length $\ell' \sim 1\mu m$ and
choosing parameters assuming an experiment using a GaAs
heterostructure \cite{Wire}, the energy spacing $\delta E$ close to the
Fermi level corresponds to roughly $0.5$K. The temperature that
sets the upper limit
for the validity of our theory is the effective Kondo temperature $T_K$, with expression
$T_K =E_C^{\ast}\, \mbox{exp}(-1/2t\nu)$ in the limit $g\ell < 2t$ \cite{FN}. 
Here $E_C^{\ast} = E_C(1
-4(t \nu)^{2}+ \cdots )$
is the renormalized charging energy \cite{WEG}, and $\nu$ is the
density of states at the Fermi level.
With $t \sim 0.2/\nu$ and $E_C \sim e^2/2C^{\phdagger}_{\Sigma}$, where $C^{\phdagger}_{\Sigma} \sim 30$aF
in a typical device, we obtain
$T_K \sim 2$K. With these estimates, our prediction in
(\ref{temperaturescaling}) applies for temperatures in the interval
$0.5$K\,$ < T < 2$K.

Considering the narrowness of the estimated temperature interval, it may experimentally be easier
to study the scaling of the capacitance with gate voltage at a fixed
temperature. Approximating the window $0.5$K\,$< T < 2$K by the $T
\rightarrow 0$ limit, the scaling can be obtained via 
a {\em Wegner expansion} \cite{wegner} of the effective (''Kondo language'') impurity free energy.
Close to the critical point $T=0, h=0$ we thus write
\begin{equation}  \label{wexp}
F_{imp}= \mbox{const.} + Tf(\frac{h}{T^{\Delta}})+g'T^{1-\Delta'}f'(\frac{
h}{T^{\Delta}})+\cdots \, .
\end{equation}
Here $f$ is a scaling function, $\Delta = 1/2$ is the boundary dimension acquired by the {\em local} magnetic 
field $h$, and $f'$ is the gradient of $f$ w.r.t. the leading irrelevant 
scaling field $g'$. The corresponding operator $\epsilon\!\otimes\!\mathcal{O}_{c}$ is 
generated from the OPE of $J^{z}$ with
$J^{z}\otimes\epsilon\otimes\mathcal{O}_{c}$ and $g'$ is thus proportional to $h$
and carries RG eigenvalue $\Delta' = -(1/K_{c}-1)/2<0$. 
In the limit $s \rightarrow \infty$, $f(s) \sim s^{1/\Delta}$. Thus, when
$T \rightarrow 0$, the second term in Eq. (\ref{wexp}) gives an
analytic contribution $\sim h^{2}$. Inspection of the third term in (10) reveals
that it can contribute a {\em finite} correction $\delta F_{imp}$ only via a term 
$\sim s^{(1-\Delta')/\Delta}$ in the expansion of $f'$, implying
that $\delta F_{imp} \sim h^{1+(1-\Delta')/\Delta}=h^{2+1/K_{c}}$. 
Contributions from higher order terms in Eq. (\ref{wexp}) are of O[$h^{4}$].
Summarizing, we obtain 
\begin{equation} \label{voltagescaling}
c(T\!=\!0,u)= D + E[K_{c}]u^{1/K_{c}}+ Fu^2 + ....
\end{equation} 
Here $D, E[K_{c}]$ and $F$ are constants, with $E[K_{c}] \rightarrow 0$ as $K_{c} \rightarrow 1$.

Before concluding, we must address the question how the boundaries of the box, as well as 
the finite range and the anisotropy of the screened Coulomb interaction, influence the physics.
Although these features must be accounted for in a faithful modeling of an experimental sample,
they will not {\em qualitatively} change the charge fluctuation effects derived in 
Eqs.~(\ref{temperaturescaling}) and (\ref{voltagescaling}):
As to the boundary effects from the quantum box, these will suppress the spectral
weight at the Fermi level, at low energies reducing the effective value of $K_{c}$ \cite{MEJ}.
The finite range $R$ of the screened Coulomb interaction further depresses $K_{c}$ by a factor
$(\mbox{ln}(R/d))^{-1/2}$, where $d$ is the (common) transverse width of wire and box \cite{Schulz}
(with $3<R/d<15$ in typical experiments on gated GaAs heterostructures \cite{Wire}).
Both effects are moderate, though, and as long as the renormalized $K_{c}$ is larger than $1/2$
the nonanalytic terms in (\ref{temperaturescaling}) and (\ref{voltagescaling}) will
remain the leading ones. Turning to the expected anisotropy $\hat{U}_{0 1}(0) \equiv g' < \hat{U}_{0 0}(0) = g$,
this will generate an exactly marginal term proportional to $(g\!-\!g')J_L^zJ_R^z$, in addition to shift the velocities in (\ref{bulkSugawara}).
While the boundary operators identified above will still be present (with $K_{c}$ renormalized upwards,
with a new perturbative expression $K_c = (1 + 2[g\!+\!g']/2)^{-1/2}$), it is
conceivable that the spin sector may now contribute additional boundary operators with noninteger
dimensions. However, if these results an exponent {\em smaller} than
$1/K_{c}$, this implies only that the
nonanalytic scaling of the capacitance gets enhanced. Conversely, the $1/K_{c}$ scaling remains the
leading one. In either case, the picture that we have uncovered by using an SU(2) invariant description
in (\ref{bulkSugawara}) will remain valid. 

In summary, we predict, under conditions specified above, that the differential capacitance of a 
quantum box side-coupled to a quantum wire  
exhibits a nonanalytic scaling in temperature and
gate voltage, with {\em the same scaling exponent in both cases}. We have traced the effect to the strong electron 
correlations inherent in one-dimensional systems, and we expect that high-precision charge measurements \cite{Dot}
should be able to detect it. An experimental verification may
shed new light on the elusive Luttinger liquid behavior of electrons in one dimension.

We thank S. Eggert, J. Kinaret and X. Wang for helpful discussions. This work was supported by 
the Swedish Research Council under grant number 621-2002-4947.


\begin{thebibliography}{99}

\bibitem{Haldane} F.~D.~M. Haldane, J. Phys. C {\bf 14}, 2585 (1981).


\bibitem{Wire} A.~R. Go$\tilde{\mbox{n}}$i \emph{et al.}, Phys. Rev. Lett. {\bf 70}, 1151 (1993);
  S. Tarucha, T. Honda, and T. Saku, Solid State
  Commun. {\bf 94}, 413 (1995); A. Yacoby \emph{et al.},
  Phys. Rev. Lett. {\bf 77}, 4612 (1996); 
  M. Rother \emph{et al.},
  Physica E {\bf 6}, 551 (2000); O.~M. Auslaender \emph{et al.},
  Science {\bf 295}, 825, (2002).

\bibitem{CNT} M. Bockrath \emph{et al.}, Science {\bf 275}, 1922
  (1997); S.~J. Tans \emph{et al.}, Nature {\bf 386}, 474 (1997).

\bibitem{Dot} 
R.~J. Schoelkopf {\em et al.}, Science {\bf 280}, 1238 (1998);
W. Lu \emph{et al.}, Nature {\bf 423}, 422 (2003).

\bibitem{Averin} D.~V. Averin and K.~K. Likharev, \emph{Mesoscopic
Phenomena in
Solids}, eds. B.~L. Altshuler, P.~A. Lee and R.~A. Webb (Elsevier Science
Publishers B.V., 1991).

\bibitem{footnote1} The case of a small quantum dot (with level spacing larger
than the temperature) is also experimentally relevant but requires a different
modelling from the one presented here for a large box.

\bibitem{Bos} A.~O. Gogolin, A.~A. Nersesyan, and A.~M. Tsvelik,
  \emph{Bosonization and Strongly Correlated Systems} (Cambridge
  University Press, 1998).

\bibitem{footnote} Note that given $\hat{U}_{0 0}(0) = \hat{U}_{1 1}(0) = \hat{U}_{01}(0) = g$, 
the absence of
renormalization of the pseudospin velocity in (\ref{bulkSugawara})
reflects the fact that there is no exchange of electrons between wire
and box away from the point contact.

\bibitem{MEJ} A.~E. Mattsson, S. Eggert, and H. Johannesson,
Phys. Rev. B {\bf 56}, 15615 (1997); F. Anfuso and S. Eggert, {\it cond-mat/0302625}.

\bibitem{Schulz} H.~J. Schulz, Phys. Rev. Lett. {\bf 71}, 1864 (1993);
W. H\"ausler, L. Kecke, and A.~H. MacDonald, Phys. Rev. B {\bf 65},
085104 (2002).

\bibitem{Mat1} K.~A. Matveev, Sov. Phys. JETP {\bf 72}(5), 892 (1991).

\bibitem{FreeDot} E. Lebanon, A. Schiller, and V. Zevin, Phys. Rev. B
{\bf
 64}, 245338 (2001); K. Le Hur and G. Seelig, Phys. Rev. B {\bf 65},
165338
(2002). 

\bibitem{InteractingDot}A. Furusaki and K.~A. Matveev, Phys. Rev. Lett. {\bf 88},
226404 (2002); E.~B. Kolomeisky, R.~M. Konik and X. Qi, Phys. Rev. B {\bf 66},
075318 (2002); E.~H. Kim, Y.~B. Kim, and C. Kallin, {\it cond-mat/0205054}. 

\bibitem{FN} A. Furusaki and N. Nagaosa, Phys. Rev. Lett. {\bf 72}, 892
  (1994).

\bibitem{BCFT} I. Affleck and A.~W.~W. Ludwig, Nucl. Phys. B {\bf 360}, 641
  (1991).

\bibitem{EK} R. Egger and A. Komnik, Phys. Rev. B {\bf 57}, 10620
(1998).

\bibitem{FJ} P. Fr\"ojdh and H. Johannesson, Phys. Rev. Lett.
75, 300 (1995); Phys. Rev. B {\bf 53}, 3211 (1996).

\bibitem{Anisotropy} I. Affleck, A.~W.~W. Ludwig, H.-B. Pang, and D.L.
Cox, Phys. Rev. B {\bf 45}, 7918 (1992);
P. Schlottmann, Phys. Rev. Lett. {\bf 84}, 1559 (2000).

\bibitem{KJ} P. Kakashvili and H. Johannesson, unpublished.

\bibitem{Field} 
  G. Zar\'and, T. Costi, A. Jerez, and N. Andrei,
  Phys. Rev. B {\bf 65}, 134416 (2002).

\bibitem{WEG} X. Wang, R. Egger and H. Grabert, Europhys. Lett. {\bf 38}, 545 (1997).

\bibitem{wegner} F.~J. Wegner, Phys. Rev. B {\bf 5}, 4529 (1972).

\end{thebibliography}
\end{document}